\documentclass[12pt]{article}
\usepackage{graphicx}
\usepackage{amsfonts}

\def\beq#1{\begin{equation}\label{#1}}
\def\eeq{\end{equation}}
\def\ber#1{\begin{eqnarray}\label{#1} \nqq}
\def\eer{\end{eqnarray}}
\newcommand{\bear}[1]{\begin{eqnarray}\label{#1}}
\newcommand{\ear}{\end{eqnarray}}

\catcode`\@=11 \@addtoreset{equation}{section}\catcode`\@=12

\newcommand{\R}{{\mathbb R}}

\newcommand{\p}{\partial}
\newcommand{\fnm}{\footnotemark}
\newcommand{\fnt}{\footnotetext}

\newcommand{\nn}{\nonumber}

\author{
  M.E. Abishev$^{1,2}$, K.A. Boshkayev$^{1,2}$,
  V.D. Dzhunushaliev$^{1,2}$ \\
  and V.D. Ivashchuk$^{3,4}$    \\
      $^{1}$ IETP, Al-Farabi KazNU, Almaty, 050040, Kazakhstan \\
   $^{2}$ Physical and Technical Faculty, Al-Farabi Kazakh National University, \\
   al'-Farabi street, 71, Almaty 050040, Kazakhstan \\
  $^{3}$ Center for Gravitation and Fundamental Metrology, VNIIMS, \\
   Ozyornaya St., 46, Moscow 119361, Russia, \\
   $^{4}$ Institute of Gravitation and Cosmology, \\
   Peoples' Friendship  University of Russia, \\
  Miklukho-Maklaya St.,6, Moscow 117198,  Russia
}

\title{Dilatonic dyon black hole solutions}

\begin{document}

\maketitle

\begin{abstract}
Dilatonic black hole dyon solutions with arbitrary dilatonic coupling constant
$\lambda \neq 0$ and canonical sign $\varepsilon = +1$ for scalar field kynetic term
are considered. These solutions are defined up to solutions of two master
equations for moduli funtions.
For $\lambda^2 \neq 1/2$ the solutions are extended to $\varepsilon = \pm 1$,
where $\varepsilon = -1$ corresponds to  ghost (phantom) scalar field.
Some physical parameters of the solutions:
gravitational mass, scalar charge, Hawking temperature, black hole area entropy and
parametrized post-Newtonian  (PPN) parameters $\beta$ and $\gamma$ are obtained.
It is shown that PPN parameters  do not depend on scalar field coupling $\lambda$ and
$\varepsilon$. Two group of bounds on gravitational mass
and scalar charge (for fixed and arbitrary extremality parameter $\mu >0$) are found by using
a certain conjecture on parameters of solutions when $1 +2 \lambda^2 \varepsilon > 0$.
These bounds  are verified numerically for certain examples.
By product we are led to  well-known lower bound on mass which was obtained
earlier by Gibbons, Kastor, London,  Townsend and Traschen by using spinor techniques.

\end{abstract}

\pagebreak

\normalsize

\section{Introduction}

At present there exists an interest to spherically-symmetric solutions, e.g. black hole and
black brane ones, related to Lie algebras and
Toda chains, see \cite{BronShikin}-\cite{GKO}
and references therein. These solutions appear
in gravitational models with scalar fields and antisymmetric forms.
Meanwhile special subclasses of low-dimensional (e.g. 4-dimensional) solutions
were not considered in detail.

Here we consider a subclass of 4-dimensional
dilatonic black hole solutions with  electric and magnetic
charges.  We  extend dilatonic black hole dyon solution from  \cite{FIMS}
to more general case, when  dilatonic scalar field may be a ghost (or phantom) one.
The ghost field   appears in the action with a kinetic term of the
``wrong sign''. This
implies the violation of the null energy condition $p + \rho \geq
0$. At the quantum level, such fields could form a
``ghost condensate'', which would be responsible for modified
gravity laws in the infra-red limit \cite{ArkH}. Present
observational data do not exclude this possiblity, and moreover
under certain conditions the phantom scenario has a preference  \cite{Kom}.

The main goal of our paper is the search of relations on physical parameters of dyonic black holes, e.g.
bounds on gravitational mass $M$ and scalar charge $Q_{\varphi}$. This problem is
 solved here up to a conjecture, which states one to one (smooth) correspondence between
the pair  $(Q_1^2,Q_2^2)$, where $Q_1$ is electric charge and $Q_2$ is magnetic charge, and the pair $(P_1,P_2)$,
where  $P_1 > 0$ and  $P_2 > 0$ are parameters of the solutions.
This conjecture is believed to be valid for all $\lambda \neq 0$
in the case of ordinary scalar field and for $0< \lambda^2 < 1/2$ for the case of phantom scalar field.
Here we verify the bounds by using  numerical calculations.

\section{Black hole dyon solutions}

\subsection{Dyonic solutions}

We consider a model governed by the action
\bear{i.1}
 S= \frac{1}{16 \pi G}  \int d^4 x \sqrt{|g|}\biggl\{ R[g] -
 \varepsilon g^{\mu \nu}\p_{\mu} \varphi  \p_{\nu} \varphi
 \nn
 - \frac{1}{2} \exp(2\lambda \varphi)F_{\mu \nu} F^{\mu \nu} \biggr\},
\ear
where $g= g_{\mu \nu}(x)dx^{\mu} \otimes dx^{\nu}$ is  metric,
 $\varphi$ is  scalar field, $F = dA
 =  \frac{1}{2} F_{\mu \nu} dx^{\mu} \wedge dx^{\nu}$
is Maxwell $2$-form, $A = A_{\mu} dx^{\mu}$, $\varepsilon = \pm 1$,
$G$ is gravitational constant,
 $\lambda \neq 0$ is a  coupling constant and
 $|g| =   |\det (g_{\mu \nu})|$. Here we put $\lambda^2 \neq 1/2$
 for $\varepsilon = - 1$.

 Let us consider a family of dyonic black hole
solutions to the field equations corresponding to the action
(\ref{i.1}). These solutions are defined on the manifold
\beq{i.2}
 M =    (2\mu, + \infty)  \times S^2 \times  \R,
\eeq
and have the following form
\bear{i.3}
 &&ds^2 = g_{\mu \nu} dx^{\mu} dx^{\nu}
 = (H_1 H_2)^h
 \biggl\{ -  (H_1 H_2)^{-2 h}
 \left( 1 - \frac{2\mu}{R} \right)
 dt^2 \\ \nonumber
 && \qquad +  \frac{dR^2}{1 - \frac{2\mu}{R}} + R^2  d \Omega^2_{2}
  \biggr\},
 \\  \label{i.3a}
 &&\exp(\varphi)=
 \left( \frac{H_1}{H_2} \right)^{h \lambda \varepsilon},
 \\  \label{i.3b}
 &&F=
 \frac{Q_1}{R^2}   H_{1}^{-2} H_{2}^{-a} dt \wedge dR
  + Q_2 \tau.
\ear
Here  $Q_1$and $Q_2$ are charges - electric and magnetic, respectively,
 $\mu > 0$ is extremality parameter,
 $d \Omega^2_{2} = d \theta^2 + \sin^2 \theta d \phi^2$
is canonical metric on the unit sphere $S^2$
 ($0< \theta < \pi$, $0< \phi < 2 \pi$),
 $\tau = \sin \theta d \theta \wedge d \phi$
is standard volume form on $S^2$,
\beq{i.16}
 h = \frac{2}{1 + 2 \lambda^2 \varepsilon},
\eeq
and
\beq{i.16a}
  a = 2 \frac{1 - 2 \lambda^2 \varepsilon }{1 + 2 \lambda^2 \varepsilon }.
\eeq
Functions $H_s > 0$ obey the equations
\beq{i3.1}
 R^2 \frac{d}{dR} \left( R^2
  \frac{\left(1 - \frac{2 \mu}{R}\right)}{H_s}
  \frac{d H_s}{dR} \right) = - h^{-1}  Q_s^2
  \prod_{l = 1,2}  H_{l}^{- A_{s l}},
\eeq
with the following boundary conditions imposed
\beq{i3.1a}
  H_s  \to H_{s0} > 0,
\eeq
for $R \to 2\mu $, and
\beq{i3.1b}
  H_s    \to 1,
\eeq
for $R \to +\infty$, $s = 1,2$.

In (\ref{i3.1}) we denote
\beq{i.18}
    \left(A_{ss'}\right)=
  \left( \begin{array}{*{6}{c}}
     2 & a\\
     a & 2\\
\end{array}
\right) .
\eeq
For the case $\varepsilon = 1$ these solutions were presented earlier in \cite{FIMS}.
They may be obtained by using  general formulae for non-extremal (intersecting)  black brane solutions
from \cite{IMp1,IMp2,IMp3,IMtop}.

First boundary condition (\ref{i3.1a}) guarantees (up to a possible additional demand on analicity
of $H_s(R)$ in the vicinity of $R=   2 \mu$)
the existence of (regular) horizon at  $R=   2 \mu$ for the metric (\ref{i.3}).
Second condition (\ref{i3.1b}) ensures  an asymptotical (for $R \to
+\infty$) flatness of the metric.

Equations (\ref{i3.1}) may be rewritten in the following form
\beq{i2.1}
  \frac{d}{dz} \left[
  \left(1 -z\right) \frac{d y_s}{dz} \right] =
          - b  q_s^2 \exp(-2 y_s - a y_{\bar{s}} ),
\eeq
$s = 1,2$ . Here and in what follows we use the following notations:
$y_s= \ln H_s$,  $z = 2 \mu/R$, $q_s = Q_s/(2\mu)$,  $b= h^{-1}$ and
 $\bar{s} = 2,1$ for $s = 1,2$, respectively.
We are seeking solutions to equations (\ref{i2.1}) for  $z \in (0,1)$
obeying
 \bear{i2.1b}
   y_s(0) = 0, \\ \label{i2.1c}
   y_s(1) = y_{s0},
 \ear
where $y_{s0}= \ln H_{s0}$ are finite (real) numbers, $s = 1,2$.
Here $z=0$ (or, more precisely $z=+ 0$) corresponds to infinity ($R = + \infty$), while
$z=1$ (or, more rigorously, $z=1-0$ ) corresponds to the horizon ($R = 2 \mu$).

Equations (\ref{i2.1}) with the finitness conditions on the horizon (\ref{i2.1c}) imposed
 imply the following  integral of motion:
  \bear{i2.1d}
 (1 -z) \left[ \left(\frac{d y_1}{dz} \right)^2 + \left(\frac{d y_2}{dz}\right)^2
         + a \frac{d y_1}{dz}\frac {d y_2}{dz} \right]  \qquad
    +  \frac{d y_1}{dz} + \frac {d y_2}{dz}      \\ \nonumber
           - b  q_1^2 \exp(-2 y_1 - a y_{2})
           - b  q_2^2 \exp(-2 y_2 - a y_{1}) = 0.
 \ear
Equations (\ref{i2.1}) and (\ref{i2.1c}) appear for  special
solutions to Toda-type equations  \cite{IMp2,IMp3,IMtop}
\beq{i2.1T}
  \frac{d^2 z_s}{du^2}   =    b  Q_s^2 \exp(2 z_s + a z_{\bar{s}} ),
 \eeq
for  functions $z_s(u) = - y_s -   \mu b u $, $s = 1,2$, depending
on harmonic radial variable $u$: $\exp(- 2 \mu u) = 1 -z$,
  with  the following asymptotical
 behaviour for $u \to + \infty$ (on the horizon) imposed:
  \beq{i2.1as}
  z_s(u) = - \mu b u + z_{s0} + o(1),
  \eeq
  where $z_{s0}$ are constants,  $s = 1,2$.
   The  energy
 integral of motion for (\ref{i2.1T}), which is compatible with the asymptotic
 conditions (\ref{i2.1as}),
 \bear{i2.1ET}
   E = \frac{h}{2} \left[ \left(\frac{d z_1}{du} \right)^2 + \left(\frac{d z_2}{du}\right)^2
            + a \frac{d z_1}{du} \frac{d z_2}{du} \right]
            \\ \nonumber
     - \sum_{s=1,2} \frac{1}{2} Q_s^2 \exp(2 z_s + a z_{\bar{s}})  = \mu^2,
  \ear
   leads us to relation  (\ref{i2.1d}).

 {\bf Remark 1.} {\em Here we exclude the case $\lambda^2=1/2$ for $\varepsilon=-1$ from our consideration,
 since we deal  with the finite value of the parameter $h$.
But nevertheless, one can also obtain a sensible solution with a horizon for this peculiar case.
This may be achieved by using another choice of moduli functions: $\bar H_s =  H_s^{h}$ instead of
$H_s$. The (implicit) solutions given by (\ref{i.3}), (\ref{i.3a}), (\ref{i.3b}) and the master
 equations (\ref{i3.1}), rewritten in terms of new moduli functions $\bar H_s$,
 have a  sensible limit for $\lambda^2=1/2$ and $\varepsilon=-1$.
 This special case  may be a subject of a separate publication.}

\section{Some integrable cases}

At present it seems impossible to find  explicit solutions
to the equations (\ref{i3.1}), (\ref{i3.1a}), (\ref{i3.1b}) analytically.
One may try to seek the solutions in the form

\beq{i3.12}
  H_{s} = 1 + \sum_{k = 1}^{\infty} P_s^{(k)}
  \left(\frac{1}{R}\right)^k,
\eeq
where $P_s^{(k)}$ are constants, $k = 1,2,\ldots $ and
$s =1,2$.

{\bf Remark 2.} {\em The $1/R$ expansion  is widely used in
gravitational physics sometimes without any indication whether the
series like (\ref{i3.12}) is i) convergent, or ii) asymptotical
one. Here the first possibility i)  follows from analytical behaviour
of functions   $H_{s}$ with respect to $z = 2 \mu/R$ in the
vicinity of the point $z =0$, which is based on  equations (\ref{i2.1d})
and the non-degeneracy condition for the function $f(z) = 1 - z$ at $z =0$, i.e. $f(0) =
1 \neq 0$. In what follows any of two assumptions: i) or ii) 
(or even more modest one: $H_{s} = 1 +  P_s/R + o(1/R)$, for $R \to + \infty$)
is enough for our  analysis. }

Meanwhile,  there exist at least two integrable configurations
related to Lie algebras $A_1 + A_1$ and $A_2$.

\subsection{$(A_1 + A_1)$-case}

Let
\beq{i4.1}
  \lambda^2 = \frac{1}{2}, \qquad \varepsilon =1.
\eeq
This value of dilatonic coupling corresponds to string induced
model. We get $h = 1$, $a = 0$ and hence (\ref{i.18}) is the
Cartan matrix for the Lie algebra $A_1 + A_1$ ($A_1 = sl(2)$).
In this case
\beq{i4.2}
 H_s = 1 + \frac{P_s}{R},
\eeq
where
\beq{i4.3}
 P_s (P_s + 2 \mu) = Q_s^2,
\eeq
$s = 1,2$. For   positive roots
of (\ref{i4.3})
\beq{i4.3p}
    P_s  = P_{s+} =  - \mu + \sqrt{\mu^2 + Q^2_s},
\eeq
we are led to a well-defined for $R > 2\mu$ solution with asymptotically
flat metric  and  horizon at $R = 2 \mu$. We note that $(A_1 + A_1)$-dyon
solution was considered earlier in \cite{GM,ChHsuL},
see also \cite{Br0,Kyr} for  certain extensions.

\subsection{$A_2$-case}

Let

\beq{i4.4}
  \lambda^2 = 3/2, \qquad \varepsilon =1.
\eeq
This value of dilatonic coupling constant appears after reduction to four dimensions of 5-dimensional
Kaluza-Klein model. We get $h = 1/2$, $a = -1$ and (\ref{i.18}) is the
Cartan matrix for the Lie algebra $A_2 = sl(3)$.
In this case we obtain \cite{IMp2}

\beq{i4.5}
H_s = 1 + \frac{P_s}{R} + \frac{P_s^{(2)}}{R^2},
\eeq
where
\bear{i4.6}
 2 Q_s^2 = \frac{P_s (P_s + 2 \mu) (P_s + 4 \mu)}{P_1 + P_2 + 4 \mu},
 \\ \label{i4.6a}
 P_s^{(2)} = \frac{P_s (P_s + 2 \mu) P_{\bar{s}}}{2 (P_1 + P_2 + 4 \mu)},
\ear
$s = 1,2$ ($\bar{s} = 2,1$). The Kaluza-Klein uplift to $D=5$ gives us
the well-known Gibbons-Wilthire solution  \cite{GibW},
which is in an agreement with the general spherically-symmetric
dyon solution (related to $A_2$ Toda chain) from  \cite{Lee}.

\subsection{Special solution with equal charges}

There exists also a special solution
\beq{i4.7}
 H_s = \left(1 + \frac{P}{R}\right)^{b},
\eeq
with equal charges $Q_s =Q$, $s = 1,2$, satisfying
\beq{i4.8}
   Q^2 = P (P + 2 \mu).
\eeq
 We remind that $b = h^{-1}$. For   positive root
of (\ref{i4.8})
\beq{i4.8p}
    P  = P_{+} =  - \mu + \sqrt{\mu^2 + Q^2},
\eeq
we get for $R > 2\mu$ a well-defined  solution with asymptotically
flat metric  and  horizon at $R = 2\mu$.

This solution is a special case of more general
``block orthogonal'' black brane  solutions  \cite{Br,IMJ2}.
Here the power in (\ref{i4.7}) appears due to relation
\beq{i3.11b}
  b = 2 \sum_{l = 1,2} A^{s l} ,
\eeq
$s = 1,2$, where $(A^{s l}) = (A_{s l})^{-1}$.
This power is integer for  $A_1 + A_1$ and $A_2$ cases.

It should be noted that this special solution is valid for
both signes $\varepsilon = \pm 1$ and has a well-defined limit
 for $\lambda^2 = 1/2$, $\varepsilon= -1$ in agreement with
{\bf Remark 1} (here $\bar H_s = 1 +  P/R$).

\subsection{The limiting $A_1$-case}

In what follows we will use two limiting solutions: electric one
with $Q_1 = Q \neq 0$ and   $Q_2 = 0$,
\bear{i4.e}
H_1 = 1 + \frac{P}{R}, \qquad H_2 = 1,
\ear
and magnetic one  with $Q_1 = 0$ and $Q_2 = Q \neq 0$,
 \bear{i4.m}
H_1 = 1, \qquad  H_2 = 1 + \frac{P}{R}.
 \ear
In both cases $P  =   - \mu + \sqrt{\mu^2 + b Q^2}$.
These solutions correspond to  the Lie algebra $A_1$.
In various notations the solution  (\ref{i4.e}) appeared
earlier in \cite{BronShikin}
and \cite{Hein,GM}, and was extended to multidimensional
case in  \cite{Hein,GM,BBFM,BI}.\fnm[1]\fnt[1]{The results of \cite{GM} seems to be correct ones
up to a typo in the first formula (2.1) for the action in \cite{GM} which should be eleminated:
the kinetic term for the scalar field should be multiplied
by extra factor $1/2$.}  A special case with $\lambda^2 = 1/2$, $\varepsilon= 1$.
was considered earlier in \cite{Gibbons,GHS}.

\section{Physical parameters}

Here we consider certain physical parameters
corresponding to the solutions under consideration

\subsection{Gravitational mass and scalar charge}

For (ADM) gravitational mass we get from (\ref{i.3})
\beq{i5.1}
 GM =   \mu +  \frac{h}{2} (P_1 + P_2),
\eeq
where  parameters $P_s = P_s^{(1)}$ appear in
the relation (\ref{i3.12}) and $G$ is the gravitational constant.

The scalar charge just follows  from (\ref{i.3a})
\beq{i5.1s}
 Q_{\varphi} = \lambda h \varepsilon (P_1 - P_2).
\eeq

For the symmetric case $Q_1^2 =Q_2^2 = Q^2 = P(P+2 \mu)$  with $P >0$ we get
$P_1 = P_2= b P$ and hence

\beq{i5.1sim}
  GM =  \mu + P = \sqrt{\mu^2 + Q^2}, \qquad   Q_{\varphi} = 0.
 \eeq

In this case the gravitational mass and the scalar charge do not depend upon
 $\lambda$ and  $\varepsilon$. The mass $M$ monotonically increases from
$\mu$ (for $Q^2= +0$) to $+ \infty$ (for $Q^2= + \infty$).

For fixed charges $Q_s$ and extremality parameter $\mu$
the mass $M$ and scalar charge $Q_{\varphi}$ are not
independent but obey a certain constraint. Indeed,
for fixed parameters $P_s = P_s^{(1)}$ in decomposition  (\ref{i3.12}) we
get
\beq{i5.12}
      y_s = \ln H_s = \frac{P_s}{2\mu} z + O(z^2),
 \eeq
 for   $z \to + 0$, which after substitution into  (\ref{i2.1d}) gives us (for $z =0$)
the following identity

\beq{i5.1p}
  P_1^2 + P_2^2  + a  P_1 P_2 + 2 \mu (P_1 + P_2)  = b (Q_1^2 + Q_2^2).
\eeq

By using  relations (\ref{i5.1}) and (\ref{i5.1s}) this identity may be rewritten
in the following form

\beq{i5.1id}
     2 (GM)^2   +   \varepsilon  Q_{\varphi}^2   = Q_1^2 + Q_2^2 + 2 \mu^2.
\eeq

It is remarkable that this formula does not contain $\lambda$.
We note that in the extremal case $\mu = +0$ this relation for $\varepsilon = 1$
 was obtained earlier in \cite{PTW}.   In derivation of (\ref{i5.1id}) the following identities were
used
\beq{i5.1i}
       a + 2 = 2 h, \qquad 2-a = 4 \varepsilon \lambda^2 h.
\eeq

{\bf Remark 3.}
{\em The paper \cite{PTW} is an important one due to the following non-trivial result
which was obtained numerically:
for $\varepsilon = 1$ the global extension of the metric  (\ref{i.3}) has two horizons only
if $\lambda^2 = p(p+1)/4$, $p =1,2,...$. Recently, this rule was explained in
 \cite{GKO} in terms of  analyticity of the dilaton at the  $AdS^2 \times S^2$ event horizon.}

\subsection{The Hawking temperature and  entropy}

The Hawking temperature corresponding to
the solution is found to be
  \beq{i5.2}
 T_H=   \frac{1}{8 \pi \mu}  (H_{10} H_{20})^{- h},
 \eeq
where $H_{s0}$ are defined in (\ref{i3.1a}).
Here and in what follows we put $c= \hbar = \kappa =1$.

For the symmetric case $Q_1^2 =Q_2^2 = Q^2 = P(P+2 \mu)$  with $P >0$ we get
\beq{i5.3sim}
  T_H =  \frac{1}{8 \pi \mu} \left(1 + \frac{P}{2 \mu}\right)^{-2}.
 \eeq
We see, that in this case the Hawking temperature $T_H$ does not depend upon
the choice of $\lambda$ and  $\varepsilon$. It monotonically decreases from
$1/(8 \pi \mu)$ (for $Q^2= +0$) to $0$ (for $Q^2= + \infty$). (Here $\mu$ is fixed.)

The Bekenstein-Hawking (area) entropy $S = A/(4G)$,
corresponding to the horizon at $R = 2\mu$, where $A$ is the horizon area,  reads
\beq{i5.2s}
S_{BH} =   \frac{4 \pi \mu^2}{G}  \left(H_{10} H_{20}\right)^{h}.
\eeq
It follows from (\ref{i5.2}) and (\ref{i5.2s}) that the product

\beq{i5.2st}
  T_H  S_{BH} =   \frac{\mu}{2G}
\eeq
does not depend upon  $\lambda$,  $\varepsilon$ and charges $Q_s$.
This product does not use explicit form of the moduli functions $H_s(R)$.

\subsection{PPN parameters}

Now we introduce a new radial variable $\rho$ by the relation
 $R =   \rho (1 + (\mu/2\rho))^2$
($\rho > \mu/2$), which gives us the 3-dimensionally
conformally-flat form of the metric  (\ref{i.3})

\bear{i5.3}
g^{(4)} =   U \Biggl\{ -
U_1 \frac{\left(1 - (\mu/2\rho) \right)^2}
{\left(1 + (\mu/2\rho) \right)^2} dt \otimes dt +
\left(1 + \frac{\mu}{2 \rho} \right)^4
\delta_{ij} dx^i \otimes dx^j \Biggr\},
\ear
where  $\rho^2 =  |x|^2 =   \delta_{ij}x^i x^j$ ($i,j =   1,2,3$)
and
\beq{5.5.1a}
U =   \prod_{s = 1,2} H_s^{h},\qquad U_1 =   \prod_{s  = 1,2} H_s^{-2 h}.
\eeq

The parametrized post-Newtonian (PPN) parameters
$\beta$ and $\gamma$ are defined by following standard relations
\bear{A.1}
g^{(4)}_{00} =   - (1 -  2 V + 2 \beta V^2 ) + O(V^3),
\\
\label{A.2}
g^{(4)}_{ij} =   \delta_{ij}(1 + 2 \gamma V ) + O(V^2),
\ear
$i,j =   1,2,3$, where, $V =   GM/\rho$
is  Newton's potential, $G$ is the gravitational constant and
$M$ is the gravitational mass (in our case given by (\ref{i5.1})).

The calculations of PPN (or Edington) parameters for the metric (\ref{i5.3})
give us the same result as in \cite{FIMS}:
\beq{i5.8}
\beta  = 1 +   \frac{1}{4(GM)^2}  (Q_1^{2} + Q_2^{2}), \qquad \gamma = 1.
\eeq

These parameters  do not depend upon $\lambda$
and $\varepsilon$. They may be calculated just without knowledge of explicit
relations for  functions $H_s(R)$.

It should be noted that (at least  formally) these parameters
obey  the observational restrictions for the solar system \cite{Wil},
when the   ratious $Q_s/(2GM)$ are small enough.

\section{Bounds on  mass and scalar charge, and  numerical calculations}

Here we start with the following hypothesis which is supported by numerical calculations.

{\bf Conjecture.}
{\em For any $h>0$,  $\varepsilon = \pm 1$, $Q_1 \neq 0$, $Q_2\neq 0$ and $\mu > 0$:
A) the moduli functions $H_s(R)$, which obey  (\ref{i3.1}), (\ref{i3.1a}) and (\ref{i3.1b}),
are uniquely defined and hence the parameters $P_1$, $P_2$, the gravitational mass $M$ and
the scalar charge $Q_{\varphi}$ are uniquely defined too;
B) the parameters $P_1$, $P_2$ are positive and the  functions $P_1 = P_1(Q_1^2,Q_2^2)$,
 $P_2 = P_2(Q_1^2,Q_2^2)$  define a diffeomorphism of $\R_{+}^2$ ($\R_{+} = \{x| x>0 \}$);
  C) in the limiting case we have: (i) for $Q_2^2 \to + 0$: $P_1 \to   - \mu + \sqrt{\mu^2 + b Q_1^2}$, $P_2 \to +0$
  and (ii) for $Q_1^2 \to + 0$: $P_1 \to +0$, $P_2 \to   - \mu + \sqrt{\mu^2 + b Q_2^2}$  ($b = h^{-1}$) .}

 The {\bf Conjecture} could be readily verified  for the case  $\varepsilon = 1$,
$\lambda^2 = 1/2$. Another integrable case  $\varepsilon = 1$, $\lambda^2 = 3/2$ is more involved and it
needs some efforts in verifying  this conjecture.

It seems that the point B) is at the same time the most crucial and most difficult to prove.
This is the main part of the conjecture.

For $h>0$ we are led to the following   bounds on the gravitational mass $M$
 and scalar charge $Q_{\varphi}$ ($Q_1 \neq 0$, $Q_2\neq 0$)

\beq{i5.13p}
   \mu  + \frac{h}{2}\left(- \mu + \sqrt{b (Q_1^2 + Q_2^2) + \mu^2} \right) < GM
    \leq \sqrt{\frac{1}{2} (Q_1^2 + Q_2^2) + \mu^2},
  \eeq
for $\varepsilon = +1$ ($\lambda \neq 0$, $0< h < 2$),
\beq{i5.13m}
 \sqrt{\frac{1}{2} (Q_1^2 + Q_2^2) + \mu^2} \leq GM
       < \mu  + \frac{h}{2} \left(- \mu + \sqrt{b (Q_1^2 + Q_2^2) + \mu^2} \right) ,
  \eeq
for $\varepsilon = -1$ ($0 < \lambda^2 < \frac12$, $h>2$) and
\beq{i5.4}
 |Q_{\varphi}| < |\lambda| h \left(- \mu + \sqrt{b (Q_1^2 + Q_2^2)+ \mu^2} \right),
  \eeq
which are valid for all $\lambda \neq 0$.

We illustrate the bounds on $M$ and $Q_{\varphi}$ graphically by two figures,
which represent a  set of physical parameters  $GM$ and $Q_{\varphi}$ for
 $Q_1^2 + Q_2^2 = Q^2 =2 $ and $\mu = 1$.
Figure 1 corresponds to the case   $\varepsilon = +1$ and $\lambda= \sqrt{ \frac{1}{2}}$, while
 Figure 2 describes the limiting case $\varepsilon = -1$ and $\lambda= \sqrt{0.499}$.
The middle points of these two arcs correspond to symmetric solutions with $Q_1^2 = Q_2^2 = 1$,
while the boundary points of the arcs correspond either to $Q_1^2 = + 0$, $Q_2^2 = 2$, or
to  $Q_1^2 = 2$, $Q_2^2 = +0$.

\begin{figure}
 \centering
  \includegraphics{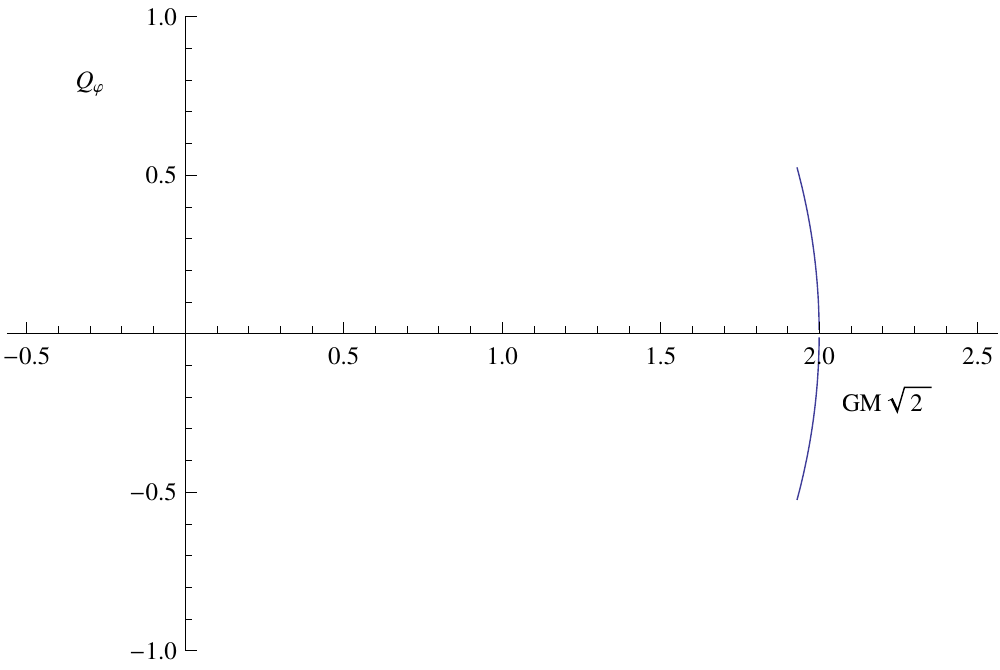}
  \caption{Graphical illustration of  bounds on $M$ and $Q_{\varphi}$ for  $\varepsilon = 1$,
  $\lambda= 1/\sqrt{2}$, $\mu = 1$ and  $Q_1^2 + Q_2^2 = 2$. }
  \label{f1.}
\end{figure}

\begin{figure}
  \centering
  \includegraphics{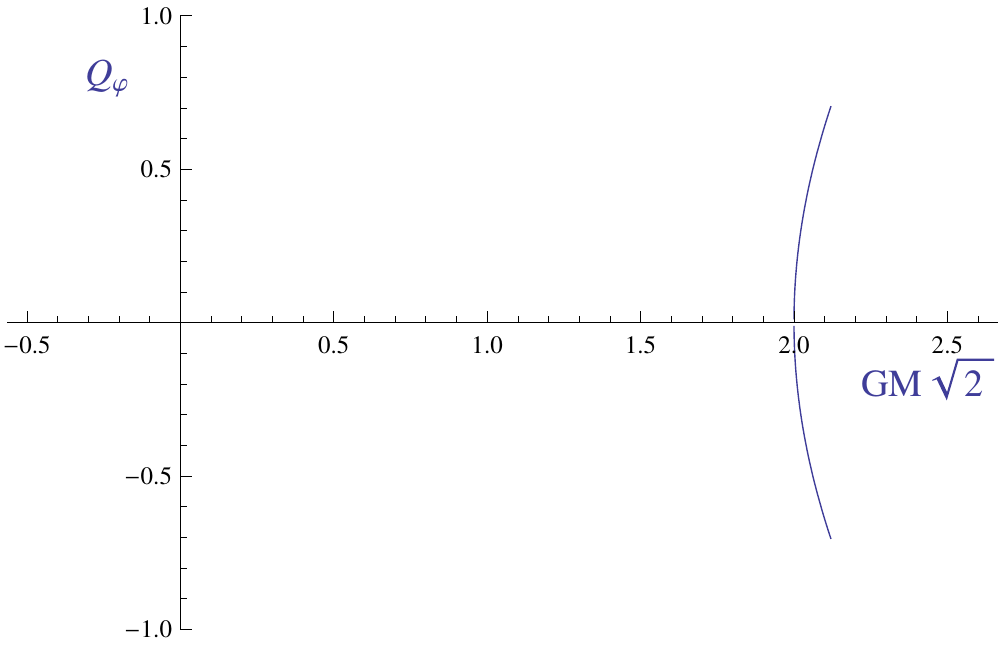}
  \caption{Graphical illustration of  bounds on $M$ and $Q_{\varphi}$ for  $\varepsilon = -1$,
   $\lambda= \sqrt{0.499 }$, $\mu = 1$ and  $Q_1^2 + Q_2^2 = 2$. }
  \label{f2.}
\end{figure}

{\bf Proof of the bounds.}
Let us prove the relations (\ref{i5.13p}), (\ref{i5.13m})  and (\ref{i5.4}) using the {\bf Conjecture}.
The right inequality (or equality) in (\ref{i5.13p}) just follows
from the relation  (\ref{i5.1id}), while the left inequality (or equality) in (\ref{i5.13m}) follows
from   (\ref{i5.1id}), and $M > 0$ which is valid due to relation (\ref{i5.1}), $h>0$ and inequalities
$P_1 >0$, $P_2 > 0$ (due to {\bf Conjecture.}).
Now let us verify the left inequality in (\ref{i5.13p}). We fix the charges by the relation
$Q_1^2 + Q_2^2 = Q^2$, $Q >0$, and put $Q_1^2 = \frac{1}{2} Q^2(1+ x)$, $Q_2^2 = \frac{1}{2} Q^2(1 - x)$, where
$-1 < x < 1$. Due to (\ref{i5.1id}) and $M > 0$ we can use the following parametrization
   \beq{i5.14}
         \sqrt{2}GM = R\cos{\psi}, \quad  Q_{\varphi} = R \sin{\psi}, \qquad   R = \sqrt{Q^2 + 2 \mu^2},
     \eeq
where $|\psi| < \pi/2$. Owing  to {\bf Conjecture} and relations (\ref{i5.1}), (\ref{i5.1s}) we get that $\psi = \psi(x)$
is a smooth function which obey
    \beq{i5.15}
        \psi(0) = 0, \qquad \psi( \pm 1 \mp 0) = \pm \psi_0.
      \eeq
Here  $R \cos{\psi_0} = \sqrt{2} ( \mu +  \frac{h}{2} P )$ and $R \sin{\psi_0} = \lambda h P $, where
 $P = -\mu +  \sqrt{b Q^2+ \mu^2}$. The limit $x \to +1 -0$ corresponds to pure electric black hole while the limit
 $x \to - 1 +0$ corresponds to pure magnetic one. To prove the relations (\ref{i5.13p})   and (\ref{i5.4})
 one should verify the inequality
     \beq{i5.15a}
        - \psi_0 < \psi(x) <  \psi_0
      \eeq
for all $x \in (-1,1)$.  Let us suppose that (\ref{i5.15a}) is not valid. Without loss of generality we
 put   $\psi(x_{*}) \geq \psi_0$   for some $x_{*}$. Then  using  (\ref{i5.15}) and the smoothness of
the function $\psi(x)$ we get that for some $x_1 \neq x_2$:  $\psi(x_{1}) = \psi(x_{2})$.
\fnm[2]\fnt[2]{This may be readily proved by using Intermediate Value Theorem. This theorem states that
if  $f(x)$ is a continuous function on the interval $[a, b]$ then for any  $d \in [f (a), f (b)]$,  there is a point
$c \in   [a, b]$ such that $f(c) = d$. (Here for $f(a) > f(b)$, $[f (a), f (b)]$ is meant as $[f (b), f (a)]$.)}
Hence for two different
sets  $(Q_1^2,Q_2^2)_1 \neq (Q_1^2,Q_2^2)_2$ we obtain the same coinciding sets: $(GM,Q_{\varphi})_1 = (GM,Q_{\varphi})_2$
and  $(P_1,P_2)_1 = (P_1,P_2)_2$, see (\ref{i5.1}) and (\ref{i5.1s}). But due to our {\bf Conjecture}
the map   $(Q_1^2,Q_2^2) \mapsto (P_1,P_2)$ is bijective one. This implies   $(P_1,P_2)_1 \neq (P_1,P_2)_2$.
We get a contradiction which proves our proposition for $\varepsilon = 1$ and arbitrary $Q_1^2 + Q_2^2 > 0$.
The proof of right inequality in   (\ref{i5.13m})  and bounds  (\ref{i5.4}) for $\varepsilon = - 1$ is quite analogous one.
 The only difference here the use of parametrization
 \beq{i5.14m}
         \sqrt{2}GM = R\cosh{\psi}, \quad  Q_{\varphi} = R \sinh{\psi}, \qquad   R = \sqrt{Q^2 + 2 \mu^2},
     \eeq
instead of  (\ref{i5.14}).

The inequalities (\ref{i5.13p}), (\ref{i5.13m}) and (\ref{i5.4})
imply the following bounds on mass and scalar charge, which are valid for all
$\mu >0$ and $h>0$
\beq{i5.16p}
   \frac{1}{2}\sqrt{h (Q_1^2 + Q_2^2)}  < GM,
     \eeq
for $\varepsilon = +1$ ($\lambda \neq 0$, $0< h < 2$),
\beq{i5.16m}
 \sqrt{\frac{1}{2} (Q_1^2 + Q_2^2)} < GM,
  \eeq
for $\varepsilon = -1$ ($0 < \lambda^2 < \frac12$, $h>2$), and
\beq{i5.16s}
 |Q_{\varphi}| < |\lambda|   \sqrt{h (Q_1^2 + Q_2^2)}
\eeq
for both cases.

The bound (\ref{i5.16p}) is in agreement with  the bound (6.16) from ref. \cite{GKLTT} (BPS-like inequality), which was proved there
 by using certain  spinor techniques.

It should be noted that in the pure  electric case for $\varepsilon = +1$ Gibbons and Wells have proved another bound
which relates $M$, $Q_{\varphi}$,  $|Q_1|$ and $\lambda$ \cite{GW}. An open question is whether such bound could be
extended somehow to the dyonic case.

{\bf Remark 4.} {\em For $h < 0$ the {\bf Conjecture} is not valid. This may be verified just by analysing the solutions
with small charge $Q_1$ (or $Q_2$).}

{\bf Remark 5.}{\em It should be noted that here we are dealing with a very special class of solutions with phantom scalar field
($\varepsilon = -1$).
Even in the  limiting case  $Q_2 = +0$ and $Q_1 \neq 0$ there exist several
branches of phantom black hole solutions which are not covered by our analysis \cite{CFR} (see also \cite{ACFR}.)}

The bounds (\ref{i5.13p}), (\ref{i5.13m}) and  (\ref{i5.4}) could be verified numerically by using
the prescription  which is described below.
We outline some results of numerical calculations which are based
on dynamical equations  (\ref{i2.1}). We start with putting the boundary conditions
on the horizon $z= 1$: $y_s(1)$, $s = 1,2$.

Then for the first derivatives on the horizon
$(\frac{d y_s}{dz})_{|z=1} = y_s^{\prime}(1)$ we
obtain from (\ref{i2.1})
 \beq{i5.9}
      y_s^{\prime}(1) = b q_s^2 \exp(-2 y_{s}(1) - a y_{\bar{s}}(1)),
 \eeq
$s = 1,2$ ($\bar{s}= 2,1$). For practical calculations we put $z= 1 - \delta$, where $\delta $ is small
enough, say $\delta = 10^{-5}$, for initial values $y_s(1)$ about $1$.
This is necessary for a correct formulation of the Cauchy problem for equations (\ref{i2.1}).

Our strategy is the following one. For fixed  $\lambda$ and $\varepsilon$ we
 start with the exact symmetric solution  obeing
  $y_1(0) = y_2(0)= 0$, i.e. we put
 \beq{i5.10}
      y_1(1) = y_2(1)  = b \ln (1+p), \qquad  q_1^2 = q_2^2 = p(p+1).
  \eeq
See (\ref{i4.7}) and (\ref{i4.8}). Here $p = P/(2\mu) > 0$.
Then we disturb relations (\ref{i5.10}) as follows

 \beq{i5.10k}
      y_1(1) =  b \ln (1+p), \quad  y_2(1)  = k b \ln (1+p), \quad  q_1^2 = q_2^2 = p(p+1),
  \eeq
  where $k \neq 1$. We get a numerical solution with $y_1(0)$ and $y_2(0)$ not obviously equal to $0$.

Now, we make a shift in our solutions
 \beq{i5.11}
     \bar{y}_s(z) = y_s(z) - y_s(0),
  \eeq
$s = 1,2$.

The functions $\bar{y}_s(z)$ give us a new solution to
Toda-like equations (\ref{i2.1}) with rescaled charges
 \beq{i5.11q}
      \bar{q}_s^2 = q_s^2 \exp(2 y_s(0) + a y_{\bar{s}}(0)),
 \eeq
$s = 1,2$. The crucial point here is that  $\bar{y}_s(z)$
obey the boundary conditions: $\bar{y}_s(0) = 0$, $s = 1,2$.

The asymptotical parameters $P_s$ are extracted from the
relations (\ref{i5.12}) (with $y_s $ replaced by $\bar{y}_s $ ).
The accuracy of calculations is controlled
by  (\ref{i2.1b}) and  (\ref{i5.1p}).

Here we present certain examples of numerical data collected in Table 1 and Table 2.
These data obey the bounds (\ref{i5.13p}), (\ref{i5.13m}) and  (\ref{i5.4}).
Of course, these tables may be enlarged by adding
(a vast number of) new lines.

\begin{table}[h]
\begin{center}
\begin{tabular}{|c|c|c|c|c|}
  \hline
  $Q_1$     & $Q_2$     & GM      & $Q_\varphi$ \ &
  bounds  on $M$ and $Q_{\varphi}$  \\\hline
  0.233313  & 0.165107  & 1.03923 & 0.0128309   & true  \\
  0.233641  & 0.182372  & 1.0421  & 0.0100296   & true\\
  0.234003  & 0.199861  & 1.04528 & 0.00693443  & true\\
  0.234398  & 0.217596  & 1.04876 & 0.00353771  & true\\
  0.234828  & 0.235605  & 1.05256 & -0.000169283  & true\\
  0.235293 & 0.253911 & 1.05669 & -0.00419616   & true\\
  \hline
\end{tabular}
\caption{Examples of numerical calculations for $\varepsilon=1; \lambda=0.5; \mu = 1.$}
\end{center}
\begin{center}
\begin{tabular}{|c|c|c|c|c|}
  \hline
  $Q_1$       & $Q_2$       & GM      & $Q_\varphi$   &
  bounds  on $M$ and $Q_{\varphi}$ \\\hline
  0.00408717  & 0.0111095  & 1.00007  & 0.0000533559  & true\\
  0.00408725  & 0.0122205  & 1.00008  & 0.0000663148   & true\\
  0.00408733  & 0.0133316  & 1.0001  & 0.0000805081   & true\\
  0.00408743  & 0.0144426  & 1.00011  & 0.0000959358   & true\\
  0.00408753  & 0.0155537  & 1.00013  & 0.000112598  & true\\
  0.00408764  & 0.0166649  & 1.00015  & 0.000130495  & true\\
  \hline
\end{tabular}
\caption{Examples of numerical calculations for $\varepsilon=-1; \lambda=0.5; \mu = 1 .$}
\end{center}
\end{table}

\section{Conclusions}

In this paper a family of non-extremal black hole dyon
 solutions in a 4-dimensional model
with a scalar field  is presented. The scalar field is
either ordinary ($\varepsilon = +1$) or ghost one ($\varepsilon = -1$).
The solutions are defined up to two functions $H_1(R)$ and
$H_2(R)$,  which obey two differential equations  of  second order
with boundary conditions imposed. For $\varepsilon = +1$ these
equations are integrable for two cases  when $\lambda^2 = 1/2$
or $\lambda^2 = 3/2$. There is also a special solutions with coinciding
electric and magnetic charges: $Q_1 =Q_2$, which is defined for all
(admissible) $\varepsilon$ and $\lambda$.

Here we have also calculated some physical parameters of the solutions:
gravitational mass $M$, scalar charge $Q_{\varphi}$, Hawking temperature,
black hole area entropy and post-Newtonian parameters  $\beta$,  $\gamma$.
We have obtained a formula which relates  $M$,  $Q_{\varphi}$,
dyon charges $Q_1$, $Q_2$, and the extremality parameter $\mu$
 for all values of  $\lambda \neq 0$. Remarkably, this formula does not contain $\lambda$.
We have also shown that the product of the Hawking temperature and
the Bekenstein-Hawking entropy does not depend upon $\varepsilon$,  $\lambda$
and the moduli functions of the solutions $H_s(R)$ as well.

We have calculated the PPN parameters  $\beta$ and $\gamma$
without knowledge of  explicit formulas for  $H_s(R)$.
The only assumption was used that these functions  
 are given (at least) by asymptotical series in $1/R$ in the  vicinity of the 
 zero point.
We have found  that $\gamma =1$ and $\beta$ does not depend upon
 $\lambda$ and $\varepsilon$.

 Here we have obtained  bounds on gravitational mass and scalar charge for
 $1 +2 \lambda^2 \varepsilon > 0$  which are based on the {\bf Conjecture} (from Section 5)
 on parameters of solutions $P_1 = P_1(Q_1^2,Q_2^2)$,  $P_2 = P_2(Q_1^2,Q_2^2)$.
 We have also presented several  results of numerical calculations which support our bounds.
 A rigorous proof of this  conjecture may be a subject of a separate publication as well a detailed consideration
 of the case $ \lambda^2 > 1/2$,  $\varepsilon  = -1$.
 For $\varepsilon  = +1$ we have also deduced from our {\bf Conjecture}  the well-known
 (unsaturated) lower bound on  mass, which was  obtained earlier by Gibbons, Kastor, London,
  Townsend and Traschen \cite{GKLTT} by using certain  spinor techniques (just like in the well-known Nester-Witten approach).

 An open question here is to find some physical (e.g. astrophysical) applications of the dyonic black hole solutions.
  Here one  may consider a possible description of  the black hole which is ``located'' at the Galactic Center.
  Recently, it was shown  that near extremal Reissner-Nordstr\"om black hole provides a better fit of recent
 observational data for the black hole  at the Galactic Center in comparison with the Schwarzschild
 black hole, see \cite{Z} and refs. therein.  Dilatonic dyon black hole solutions (with certain scalar charge) may be
 used for a search of the best fit of the observational data for the black hole at the Galactic Center.
 For such research the thermodynamical calculations may be of relevance, e.g. due to possible analysis
 of the black hole stability, seach of bounds on  variations of physical parameters etc.

\vspace{5pt}

{\bf Acknowledgment}
The paper is supported by the program of KazNU (Almaty) on 2014 year. VD acknowledges support from a grant No.~0263/PCF~--~14 in fundamental research in natural sciences by the Ministry of Education and Science of Kazakhstan.

\end{document}